# Single-shot optical readout of a quantum bit using cavity quantum electrodynamics

Shuo Sun and Edo Waks

Department of Electrical and Computer Engineering, Institute for Research in Electronics and Applied Physics, and Joint Quantum Institute, University of Maryland, College Park, Maryland 20742, USA

**Abstract**

We propose a method to perform single-shot optical readout of a quantum bit (qubit) using cavity quantum electrodynamics. We selectively couple the optical transitions associated with different qubit basis states to the cavity, and utilize the change in cavity transmissivity to generate a qubit readout signal composed of many photons. We show that this approach enables single-shot optical readout even when the qubit does not have a good cycling transition required for standard resonance fluorescence measurements. We calculate the probability that the measurement detects the correct spin state using the example of a quantum dot spin under various experimental conditions and demonstrate that it can exceed 0.99.

## I. INTRODUCTION

The ability to projectively measure the state of a qubit with high accuracy is crucial for nearly all quantum information processing applications[1]. In the majority of applications, these qubit measurements must be performed in a single shot[2]. For example, quantum computing requires the ability to read out the states of all output qubits after the quantum algorithm completes[3-5], and quantum cryptography requires readout of all transmitted qubits[6,7]. Single shot qubit readout also plays an important role in quantum error corrections[8], quantum teleportation[9], and experimental measurements of quantum non-locality[10].

Resonance fluorescence spectroscopy is currently one of the most effective ways to perform optical single-shot readout. However, this approach requires a cycling transition where an excited state optically couples to only one of the qubit basis states. The cycling transition yields a large number of resonance fluorescence photons for one basis state, enabling strong optical signal even with poor detection efficiency, while yielding very few photons for the other state. Previous studies have demonstrated single-shot readout with resonance fluorescence in a number of qubit systems, including cold Rubidium atoms[11-13], trapped Calcium[14] and Ytterbium[15] ions, and nitrogen vacancy centers[16,17]. However, this readout method does not work for qubit systems that lacks a cycling transition such as spins of singly charged quantum dots[18], trapped Aluminum ions[19], and fluorine impurities in CdTe[20]. Cavity quantum electrodynamics provides an alternative approach to optically detect the qubit. This approach utilizes the reflectivity or transmissivity of the cavity which strongly depends on the quantum state of the qubit due to their coupling[21-26]. Several works have proposed cavity enhanced single-shot qubit readout based on this principle[27-29]. However these proposals still assume that the qubit possess a good cycling transition.

In this paper, we propose a protocol for single-shot optical readout of a qubit that lacks a good cycling transition based on cavity quantum electrodynamics. We analyze the specific example of a singly charged InAs quantum dot with a magnetic field applied perpendicular to its growth direction (Voigt geometry), a particularly suitable qubit system for our approach, and show that the probability of detecting the correct spin (which we refer as success probability) can exceed 99% under realistic experimental conditions. In addition, we demonstrate that the success probability is very robust to emitter dephasing and spectral diffusion, which is particularly important for solid-state qubit implementations. Our protocol could serve as an important building block for integrated quantum circuits and on-chip quantum computation.

The paper is organized as follows. Section II describes the proposed protocol for single-shot qubit readout. In Sec. III, we derive a formalism to calculate the success probability of the qubit readout operation. In Sec. IV, we numerically calculate the success probability under various experimental conditions using the example of the quantum dot spin. In Sec. V, we analyze the effect of emitter dephasing and spectral diffusion on the success probability.

## II.     PROTOCOL FOR SINGLE-SHOT QUBIT READOUT

Figure 1(a) shows a schematic of the proposed protocol, which consists of an optically active matter qubit inside an optical cavity. We assume that the qubit system has a λ-type energy structure as shown in Fig. 1(b). The two ground states of the system form a qubit, denoted as $|g_0\rangle$ and $|g_1\rangle$. Each qubit basis state has an optical transition with an excited state, denoted by $\mu_0$ and $\mu_1$ respectively. In contrast to resonance fluorescence techniques that require a high branching ratio between two transitions, here we are primarily interested in the situation where the two optical transitions have similar dipole strength.

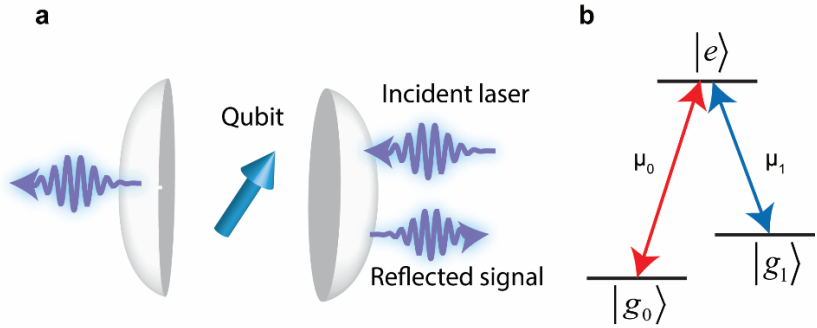

FIG. 1. (Color online) (a) Proposed scheme for optical single-shot readout of a qubit using cavity electrodynamics. (b) Energy level structure for a qubit system that lacks a cycling transition.

To perform optical readout, we consider the case where optical transition $\mu_0$ is resonant with the cavity while transition $\mu_1$ is decoupled, either by a large detuning or by selection rules if transition $\mu_1$ emits a photon with a different polarization than the cavity mode. In this configuration the coupling between the atom and cavity depends on the qubit state and will have different reflection or transmission coefficients, enabling qubit readout by optically probing the cavity. We assume that the cavity couples equally to the reflection and transmission mode.

We start the protocol by resonantly probing the cavity transmissivity using a pulsed laser of a duration $T$. We define the incident photon flux as $n^{in}$ (in units of photons per second). We assume $n^{in} \ll 1/\tau$ where $\tau$ is the modified lifetime of the excited state. Thus, the system operates in the weak excitation regime[30]. In this limit, the average number of transmitted photons is given by $N_0(T) = \frac{\eta T n^{in}}{(1+C)^2}$ and $N_1(T) = \eta T n^{in}$ respectively[30], where $N_0(T)$ and $N_1(T)$ represents the average number of collected photons when the qubit is in state $|g_0\rangle$ and $|g_1\rangle$ respectively, and $\eta$ is photon overall collection efficiency that accounts for coupling efficiency of the optics, imperfect spatial mode matching between the incident photon and the cavity, and quantum efficiency of the detector. We define the atomic cooperativity by $C = 2g^2/\kappa\gamma$, where $g$ is the coupling strength between the cavity and transition $\mu_0$, $\kappa$ is the cavity energy decay rate, and $\gamma$ is the decay rate of the excited state. Note that both the reflection and transmission ports could in principle be used to measure the spin, but we are using the transmission port here because it is less sensitive to mode-matching.

To determine the qubit state, we compare the number of collected photons with a threshold photon number $k$. When the number of collected photons is less than $k$ the measurement result reports a qubit state $|g_0\rangle$, otherwise it reports state $|g_1\rangle$. We define the probability of a successful qubit readout operation as $P_s = \max_k \{q_0 p_0(k) + q_1 p_1(k)\}$, where $q_0$ and $q_1$ is the probability that the qubit occupies state $|g_0\rangle$ and $|g_1\rangle$ respectively, and $p_0(k)$ and $p_1(k)$ is the probability of getting a correct result using threshold photon number $k$ when the qubit is initially in state $|g_0\rangle$ and $|g_1\rangle$ respectively. In Appendix A we show that the success probability is given by

$$P_s(T) = \frac{1}{2} + \frac{1}{2} \cdot \sum_{j=0}^{M} \frac{1}{j!} \cdot \left\{ [N_0(T)]^j e^{-N_0(T)} - [N_1(T)]^j e^{-N_1(T)} \right\}, \tag{1}$$

where $M = \left\lfloor \dfrac{N_1(T) - N_0(T)}{\ln[N_1(T)] - \ln[N_0(T)]} \right\rfloor$ is the threshold photon number that gives the optimal success probability, and $\lfloor x \rfloor$ indicates the largest integer that is not greater than $x$. In this derivation we have assumed $q_0 = q_1 = 0.5$, since in general one has no a-priori knowledge about the occupation probability of the two spin states.

Figure 2 shows the success probability of the qubit readout operation as a function of $T$ for several different values of the cooperativity. The success probability grows monotonically with $T$ because we collect more photons. We are able to achieve near unity success probability as long as the probe pulse duration is long enough, even for a very small cooperativity. The ripples in the plot are because the optimal threshold photon number can only increase by a discrete step of 1.

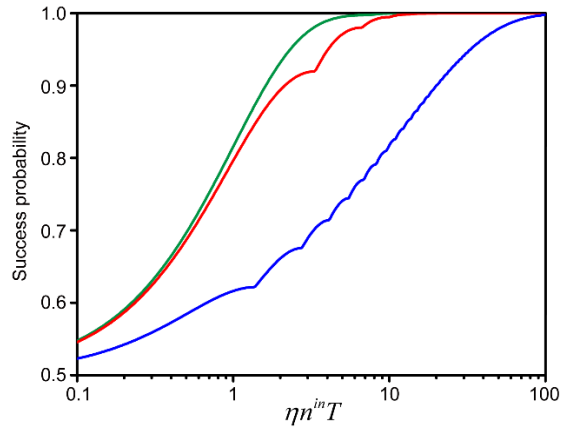

FIG. 2. (Color online) Success probability of the qubit readout operation as a function of probe pulse duration for several different values of cooperativity. Blue, $C = 0.4$; Red, $C = 4$; Green, $C = 40$.

## III.    SUCCESS PROBABILITY IN THE PRESENCE OF QUBIT FLIP

In the previous section we showed that the success probability of the qubit readout operation monotonically increases with the probe pulse duration and can eventually approach unity. However, the probe pulse duration is fundamentally limited by the laser induced qubit flip which causes measurement errors. Thus, to accurately calculate the success probability, we need a model that incorporates both cavity transmissivity modification and qubit flip errors.

Eq. (1) remains valid in the presence of qubit flips, but we need to derive the expressions for $N_0(T)$ and $N_1(T)$ which no longer holds a simple linear relationship with $T$. We can calculate $N_0(T)$ and $N_1(T)$ by integrating the photon output flux of the cavity over the time duration $T$, given by

$$N_{0(1)}(T) = \eta \int_0^T \left| Tr\left(\sqrt{\kappa}\hat{\mathbf{a}}\rho_{0(1)}(t)\right) \right|^2 dt, \quad (2)$$

where $\hat{\mathbf{a}}$ is the photon annihilation operator for the cavity mode, $\rho_{0(1)}(t)$ is the density matrix of the system at time $t$ when the qubit is in state $|g_0\rangle$ and $|g_1\rangle$ at $t=0$ respectively.

To calculate $\rho_{0(1)}(t)$, we numerically solve the system dynamics using the master equation given by $\frac{d\rho}{dt} = -\frac{i}{\hbar}\left[\hat{\mathbf{H}}, \rho\right] + \hat{\mathbf{L}}\rho$, where $\hat{\mathbf{H}}$ is the system Hamiltonian that accounts for all unitary processes, and $\hat{\mathbf{L}}$ is the Liouvillian superoperator that accounts for all non-unitary Markovian processes. We write the system Hamiltonian in a reference frame with respect to the frequency of the incident field ω, given by $\hat{\mathbf{H}} = \hat{\mathbf{H}}_0 + \hat{\mathbf{H}}_{int} + \hat{\mathbf{H}}_d$, where

$$\hat{\mathbf{H}}_0 = \hbar(\omega_c - \omega)\hat{\mathbf{a}}^\dagger\hat{\mathbf{a}} + \hbar(\omega_a - \omega)|e\rangle\langle e|, \quad (3)$$

$$\hat{\mathbf{H}}_{int} = ig\hbar\left(\hat{\mathbf{a}}|e\rangle\langle g_0| - \hat{\mathbf{a}}^\dagger|g_0\rangle\langle e|\right), \quad (4)$$

$$\hat{\mathbf{H}}_d = \hbar\sqrt{\kappa}\cdot\varepsilon\left(\hat{\mathbf{a}}^\dagger + \hat{\mathbf{a}}\right). \quad (5)$$

In Eqs. (3) - (5), $\omega_c$ is the frequency of the cavity mode, and $\omega_a$ is the frequency of transition $|g_0\rangle \leftrightarrow |e\rangle$.

The Liouvillian superoperator $\hat{\mathbf{L}}$ accounts for the decay of the cavity field and spontaneous emission of each optical transition. This operator is given by

$$\hat{\mathbf{L}} = \kappa D(\hat{\mathbf{a}}) + \gamma_0 D(|g_0\rangle\langle e|) + \gamma_1 D(|g_1\rangle\langle e|), \tag{6}$$

where $D(\hat{\mathbf{O}})\rho = \hat{\mathbf{O}}\rho\hat{\mathbf{O}}^\dagger - 1/2\hat{\mathbf{O}}^\dagger\hat{\mathbf{O}}\rho - 1/2\rho\hat{\mathbf{O}}^\dagger\hat{\mathbf{O}}$ is the general Linblad operator form for the collapse operator $\hat{\mathbf{O}}$. The parameters $\gamma_0$ and $\gamma_1$ are the spontaneous emission rates for transitions $|g_0\rangle \leftrightarrow |e\rangle$ and $|g_1\rangle \leftrightarrow |e\rangle$ respectively. The transition linewidth can also be broadened due to trion dephasing and spectral diffusion, which we will revisit in Sec. V.

## IV. ANALYSIS OF SUCCESS PROBABILITY

In this section, we perform numerical calculations on a specific case of a charged quantum dot coupled to a photonic crystal defect cavity. Figure 3(a) shows the energy level structure of the charged quantum dot in the presence of a magnetic field applied in the Voigt configuration[18] (which is the prerequisite for all-optical coherent spin manipulation[31,32]). It exhibits two ground states denoted by $|g_0\rangle$ and $|g_1\rangle$, and two excited states denoted by $|e_0\rangle$ and $|e_1\rangle$, enabling four optical transitions, which is slightly different from the three-level system depicted in Fig. 1(b). The vertical transitions ($|g_0\rangle \leftrightarrow |e_0\rangle$ and $|g_1\rangle \leftrightarrow |e_1\rangle$) and cross transitions ($|g_0\rangle \leftrightarrow |e_1\rangle$ and $|g_1\rangle \leftrightarrow |e_0\rangle$) couple to orthogonal polarization components of an optical field, denoted $V$ and $H$ respectively. Since all four possible transitions are optically allowed, this qubit system does not have a good cycling transition.

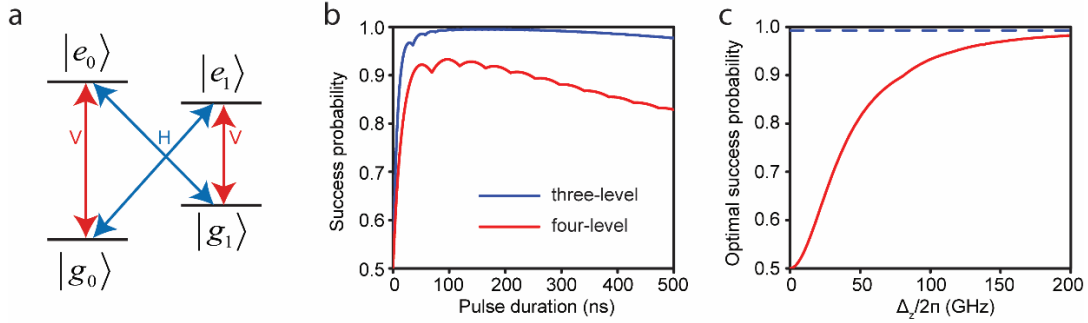

FIG. 3. (Color online) (a) Energy level structure for a charged quantum dot in the Voigt configuration. (b) Success probability of spin readout as a function of probe pulse duration calculated using the three-level model (blue solid line) and four-level model (red solid line). (c) Red solid line, optimal success probability of spin readout as a function of $\Delta_z$. Blue dashed line shows the optimal success probability calculated using a three-level model.

We assume the polarization of the cavity mode is parallel to the *V* direction, which couples it to vertical transitions $|g_0\rangle \leftrightarrow |e_0\rangle$ and $|g_1\rangle \leftrightarrow |e_1\rangle$ with equal strength $g$ and decouples it from the cross transitions due to selection rules. We consider the case where transition $|g_0\rangle \leftrightarrow |e_0\rangle$ is resonant with the cavity mode and the probe laser, whereas transition $|g_1\rangle \leftrightarrow |e_1\rangle$ is detuned by $\Delta_z$. In the limit where $\Delta_z \gg 2g^2/\kappa$, we could effectively ignore the coupling between transition $|g_1\rangle \leftrightarrow |e_1\rangle$ and the cavity, and simplify the qubit system to the model depicted in Fig. 1(b). We will revisit this assumption in the later part of this section. We set the cavity parameters to $g/2\pi = 20\,\text{GHz}$ and $\kappa/2\pi = 6\,\text{GHz}$ [33]. For the quantum dot, we assume the spontaneous emission rate is 0.1 GHz for both optical transitions ($\gamma_0/2\pi = \gamma_1/2\pi = 0.1\,\text{GHz}$)[34]. We set the incident photon field amplitude to $\varepsilon = \sqrt{0.01 \times 2g^2/\kappa}$, which corresponds to an average incident photon number of 0.01 per modified lifetime of transition $|g_0\rangle \leftrightarrow |e_0\rangle$, to ensure that we are operating in the linear weak excitation regime.

The blue curve shown in Fig. 3(b) shows the success probability as a function of laser pulse duration $T$ for a photon collection efficiency of $\eta = 1\%$. The success probability initially increases with $T$ because we collect more photons, similar as the results shown in Fig. 2. At even larger $T$ the success probability achieves a maximum and begins to decline because the probing laser induces a spin-flip. We define the optimal success probability as $P_s^{opt} = \max_T \{P_s(T)\}$, which achieves 0.995 with a time window of $T = 153$ ns.

We next investigate how a finite value of $\Delta_z$ affects the spin readout operation. In this case, we have to take into account the coupling between transition $|g_1\rangle \leftrightarrow |e_1\rangle$ and the cavity by using the four-level model (Fig. 3(a)) instead of the three-level model (Fig. 1(b)). We follow the same procedure described in Sec. III to calculate the success probability, with slight modifications of the system Hamiltonian and Liouvillian superoperator. We still write the system Hamiltonian as $\hat{\mathbf{H}} = \hat{\mathbf{H}}_0 + \hat{\mathbf{H}}_{int} + \hat{\mathbf{H}}_d$, but we modify $\hat{\mathbf{H}}_0$ and $\hat{\mathbf{H}}_{int}$ to

$$\hat{\mathbf{H}}_0 = \hbar(\omega_c - \omega)\hat{\mathbf{a}}^\dagger\hat{\mathbf{a}} + \hbar(\omega_a - \omega)|e_0\rangle\langle e_0| + \hbar(\omega_a - \Delta_z - \omega)|e_1\rangle\langle e_1|, \tag{7}$$

$$\hat{\mathbf{H}}_{int} = ig\hbar(\hat{\mathbf{a}}|e_0\rangle\langle g_0| - \hat{\mathbf{a}}^\dagger|g_0\rangle\langle e_0|) + ig\hbar(\hat{\mathbf{a}}|e_1\rangle\langle g_1| - \hat{\mathbf{a}}^\dagger|g_1\rangle\langle e_1|). \tag{8}$$

We also modify the Liouvillian superoperator as

$$\hat{\mathbf{L}} = \kappa D(\hat{\mathbf{a}}) + \gamma_0 D(|g_0\rangle\langle e_0|) + \gamma_1 D(|g_1\rangle\langle e_0|) + \gamma_2 D(|g_0\rangle\langle e_1|) + \gamma_3 D(|g_1\rangle\langle e_1|), \tag{9}$$

where $\gamma_2$ and $\gamma_3$ are the spontaneous emission rates for transitions $|g_0\rangle \leftrightarrow |e_1\rangle$ and $|g_1\rangle \leftrightarrow |e_1\rangle$ respectively. We again assume the spontaneous emission rate is 0.1 GHz for all optical transitions ($\gamma_0/2\pi = \gamma_1/2\pi = \gamma_2/2\pi = \gamma_3/2\pi = 0.1$ GHz).

The red curve in Fig. 3(b) shows the spin readout success probability as a function of laser pulse duration $T$, calculated using the four-level model. In the calculation we set $\Delta_z/2\pi = 100$ GHz, corresponding to a magnetic field of 9.4 T[26], and keep all other parameters the same as the calculations for the blue curve. The success probability calculated using the four-level model has a similar trend as a function of $T$, with an optimal success probability of 0.933. This value is lower than the value calculated using the three-level model since the cavity now couples to both spin states.

We further calculate the optimal success probability as a function of $\Delta_z$, shown as the red solid line in Fig. 3(c). When $\Delta_z$ is small, the cavity couples almost equally to two spin states, leading to a low value of success probability. The optimal success probability monotonically increases with $\Delta_z$ and approaches the value calculated using the three-level model (blue dashed line), because the coupling between transition $|g_1\rangle \leftrightarrow |e_1\rangle$ and the cavity becomes negligible.

Finally, we analyze optimal success probability as a function of the overall photon collection efficiency $\eta$. Figure 4 shows $P_s^{opt}$ as a function of $\eta$, where all other parameters are set to be the same as Fig. 3(b). The success probability increases with the collection efficiency and eventually approaches 1. At collection efficiency of 2.5% the success probability achieves a value of 0.99. This efficiency is achievable with multiple cavity structures including photonic crystals[35-38] and micro-pillars[39], and is also within the range of single photon counters[40].

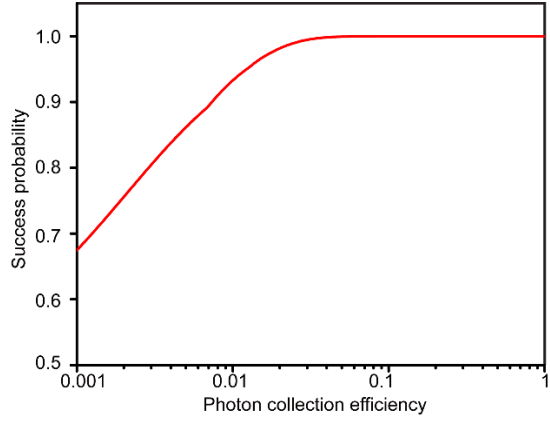

FIG. 4. (Color online) Success probability of qubit readout operation as a function of photon overall collection efficiency.

## V. QUBIT DEPHASING AND SPECTRAL DIFUSSION

To this point, we have assumed a radiatively limited linewidth for the optical transition of the qubit. In a realistic situation, the transition can be homogenously broadened due to dephasing, and inhomogeneously broadened due to spectral diffusion, which might affect the success probability of the qubit readout operation. To incorporate dephasing into our model, we introduce an additional term $\hat{\mathbf{L}}_d$ in the Liouvillian superoperator given by

$$\hat{\mathbf{L}}_d = 2\gamma_d D(|e_0\rangle\langle e_0|) + 2\gamma_d D(|e_1\rangle\langle e_1|). \tag{10}$$

The above expression assumes the same pure dephasing rate $\gamma_d$ for both excited states $|e_0\rangle$ and $|e_1\rangle$. We calculate the system dynamics using the master equation $\frac{d\rho}{dt} = -\frac{i}{\hbar}[\hat{\mathbf{H}}, \rho] + (\hat{\mathbf{L}} + \hat{\mathbf{L}}_d)\rho$, where $\hat{\mathbf{L}}$ is given by Eq. (9). We still calculate the success probability of the qubit readout operation using the example of a quantum dot spin.

The blue curve in Fig. 5 shows $P_s^{opt}$ as a function of trion dephasing rate $\gamma_d$. We assume that $\eta = 2.5\%$, which achieves success probability of 0.99 in the absence of qubit linewidth broadening.

We set all the other parameters to the same values as the ones used in Fig. 3(b). Increasing the trion dephasing rate reduces the cooperativity, which degrades the success probability. For a typical trion dephasing rate $\gamma_d/2\pi = 1\,\text{GHz}$, we are still able to achieve a success probability as high as 0.93.

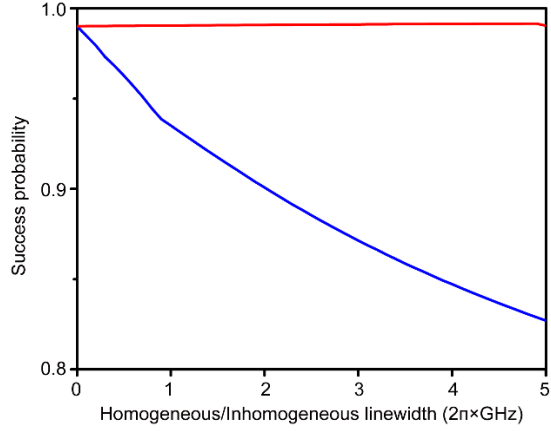

FIG. 5. (Color online) Success probability of the qubit readout operation as a function of quantum dot homogeneously broadened (blue) and inhomogeneously broadened linewidth (red).

We next consider the effect of spectral diffusion. We assume that spectral diffusion is slow compared to the laser pulse duration. We can thus model it by setting $\omega_1$ as $\omega_1 = \omega_1^{(0)} + \Delta\omega$, where $\omega_1^{(0)}$ is the average transition frequency of $|g_0\rangle \leftrightarrow |e_0\rangle$, and $\Delta\omega$ is a random variable corresponding to the frequency shift of transition $|g_0\rangle \leftrightarrow |e_0\rangle$ which may be different for each measurement shot. Here we only account for the spectral diffusion of transition $|g_0\rangle \leftrightarrow |e_0\rangle$ since the detuning between other transitions and the incident laser is much greater than their inhomogeneously broadened linewidth. We still assume the resonance condition where $\omega_c = \omega_1^{(0)} = \omega$.

Following the same procedure as described in Sec. III, we calculate the success probability $P_s(T, \Delta\omega)$ as a function of probe pulse duration $T$ for different values of $\Delta\omega$. We define the average success probability as $\bar{P}_s(T) = \int_{-\infty}^{\infty} P_s(T, \Delta\omega) G(\Delta\omega) d\Delta\omega$, where $G(\Delta\omega)$ is the probability distribution of the random variable $\Delta\omega$. Spectral diffusion is often modeled as a Gaussian distribution[41], given by

$$G(\Delta\omega) = \frac{2}{\gamma_I} \sqrt{\frac{\ln 2}{\pi}} \exp\left(-4\ln 2 \cdot \left(\frac{\Delta\omega}{\gamma_I}\right)^2\right), \tag{11}$$

where $\gamma_I$ is the inhomogeneously broadened linewidth. We define the optimal success probability as $\bar{P}_s^{opt} = \max_T \{\bar{P}_s(T)\}$.

The red curve in Fig. 5 shows $\bar{P}_s^{opt}$ as a function of inhomogeneously broadened linewidth $\gamma_I$. In this calculation we ignore the effect of trion dephasing ($\gamma_d = 0$), and we set all other parameters the same as the ones used in the Fig. 3(b). The success probability remains very robust to inhomogeneously broadened linewidth, in contrast to the case with homogenously broadened linewidth. This result might be surprising at first, since with larger inhomogeneous linewidth, the contributions from those cases where the transition $|g_0\rangle \leftrightarrow |e_0\rangle$ and the cavity are detuned become more significant. For these detuned cases, the contrast of photon collection rate between the spin-up and spin-down states are lower than the resonant case, which might degrade the success probability. To explain the robustness, we note that the detuned contributions also suppress the laser induced spin flip. The robustness of the success probability over the emitter spectral diffusion makes this protocol very appealing for experimental realizations in the quantum dot system, since spectral diffusion is the dominant term in the quantum dot linewidth in several reported experiments[22,42].

## VI. CONCLUSIONS

We have proposed a protocol for single-shot optical readout of a qubit that does not possess a cycling transition. This protocol is broadly applicable to many qubit systems including quantum dot spins[18], trapped Aluminum ions[19], and fluorine impurities in CdTe[20], and may also be useful for lots of new and emerging qubit systems that are still under developed. In particular, we have shown the feasibility of implementing this protocol on a charged quantum dot in the Voigt configuration, which could simultaneously enable all-optical coherent spin manipulation[31,32] and single-shot optical readout of the quantum dot spin, an important step towards quantum dot spin based quantum information processing. Our protocol shows how tailoring light-matter interactions opens up new possibilities for processing quantum information with higher speed and accuracy.

# ACKNOWLEDGEMENTS

The authors would like to acknowledge support from the DARPA QUINESS program (grant number W31P4Q1410003), the Physics Frontier Center at the Joint Quantum Institute, the National Science Foundation (grant number PHYS. 1415458), and the Center for Distributed Quantum Information.

# APPENDIX A: DERIVATION OF SUCCESS PROBABILITY OF THE QUBIT READOUT OPERATION

We use two random variables $X_0$ and $X_1$ to denote the number of collected photons when the qubit is initially in state $|g_0\rangle$ and $|g_1\rangle$ respectively. We assume that the dominant noise mechanism is shot noise and thus $X_0$ and $X_1$ obeys Poisson distributions with average photon number $N_0$ and $N_1$ respectively. This assumption is valid in the weak excitation regime provided the detector dead time is short compared to the photon arrival rate. In this limit, we can calculate $p_0(k)$ and $p_1(k)$ (defined in Sec. II) to be

$$p_0(k) = p(X_0 \leq k) = \sum_{j=0}^{k} \frac{N_0^j}{j!} \cdot e^{-N_0}, \tag{A1}$$

$$p_1(k) = p(X_1 > k) = 1 - \sum_{j=0}^{k} \frac{N_1^j}{j!} \cdot e^{-N_1}. \tag{A2}$$

We calculate the success probability following its definition $P_s = \max_k \{q_0 p_0(k) + q_1 p_1(k)\}$. Since in general one has no a-priori knowledge about the occupation probability of the two spin states, we assume $q_0 = q_1 = 0.5$. Substituting $p_0(k)$ and $p_1(k)$ into this definition, we obtain that

$$P_s = \frac{1}{2} + \frac{1}{2} \cdot \max_k \left\{ \sum_{j=0}^{k} \frac{1}{j!} \left( N_0^j \cdot e^{-N_0} - N_1^j \cdot e^{-N_1} \right) \right\}. \tag{A3}$$

It is straightforward to calculate that the maximum achieves at the largest integer $k$ that yields $N_0^k \cdot e^{-N_0} - N_1^k \cdot e^{-N_1} \geq 0$. This directly leads to the final results shown in Eq. (1) in Sec. II.

**Figure Captions**

FIG. 1. (Color online) (a) Proposed scheme for optical single-shot readout of a qubit using cavity electrodynamics. (b) Energy level structure for a qubit system that lacks a cycling transition.

FIG. 2. (Color online) Success probability of the qubit readout operation as a function of probe pulse duration for several different values of cooperativity. Blue, $C = 0.4$; Red, $C = 4$; Green, $C = 40$.

FIG. 3. (Color online) (a) Energy level structure for a charged quantum dot in the Voigt configuration. (b) Success probability of spin readout as a function of probe pulse duration calculated using the three-level model (blue solid line) and four-level model (red solid line). (c) Red solid line, optimal success probability of spin readout as a function of $\Delta_z$. Blue dashed line shows the optimal success probability calculated using a three-level model.

FIG. 4. (Color online) Success probability of qubit readout operation as a function of photon overall collection efficiency.

FIG. 5. (Color online) Success probability of the qubit readout operation as a function of quantum dot homogeneously broadened (blue) and inhomogeneously broadened linewidth (red).